\documentclass[conference]{IEEEtran}
\IEEEoverridecommandlockouts
\usepackage{cite}
\usepackage{graphicx}
\usepackage{amsmath,amssymb,amsfonts}
\usepackage{algorithmic}
\usepackage{underscore}
\usepackage{multirow}
\usepackage{graphicx}
\usepackage{textcomp}
\usepackage{xcolor}
\usepackage{dblfloatfix}
\def\BibTeX{{\rm B\kern-.05em{\sc i\kern-.025em b}\kern-.08em
    T\kern-.1667em\lower.7ex\hbox{E}\kern-.125emX}}

\makeatletter
\newcommand{\linebreakand}{%
  \end{@IEEEauthorhalign}
  \hfill\mbox{}\par
  \mbox{}\hfill\begin{@IEEEauthorhalign}
}
\makeatother

\begin{document}

\begin{titlepage}
\thispagestyle{empty} 
\vspace*{\fill} 
\begin{center}
    \textbf{Copyright Statement}\\[1cm]
    This paper has been accepted for publication in the IEEE CCSB 2024 conference \\

    © 2024 IEEE. Personal use of this material is permitted. Permission from IEEE must be obtained for all other uses, in any current or future media, including reprinting/republishing this material for advertising or promotional purposes, creating new collective works, for resale or redistribution to servers or lists, or reuse of any copyrighted component of this work in other works.
\end{center}
\vspace*{\fill}
\end{titlepage}

\title{Graphical Structural Learning of rs-fMRI data in Heavy Smokers\\
}


\author{
    \IEEEauthorblockN{Yiru Gong\textsuperscript{1,*}, Qimin Zhang\textsuperscript{2a}, Huili Zheng\textsuperscript{2b}, Zheyan Liu\textsuperscript{2c}, and Shaohan Chen\textsuperscript{2d}}
    \IEEEauthorblockA{
        \textsuperscript{1, 2}\textit{Department of Biostatistics, Columbia University, New York, NY 10032, USA}
    }
    \IEEEauthorblockA{
        yiru.g@columbia.edu\textsuperscript{1,*},
        qimin.zhang@columbia.edu\textsuperscript{2a},\\
        hz2710@caa.columbia.edu\textsuperscript{2b},
        zl3119@caa.columbia.edu\textsuperscript{2c}, shaohan.chen@caa.columbia.edu\textsuperscript{2d}
    }
}

\maketitle

\begin{abstract}
Recent studies revealed structural and functional brain changes in heavy smokers. However, the specific changes in topological brain connections are not well understood. We used Gaussian Undirected Graphs with the graphical lasso algorithm on rs-fMRI data from smokers and non-smokers to identify significant changes in brain connections. Our results indicate high stability in the estimated graphs and identify several brain regions significantly affected by smoking, providing valuable insights for future clinical research.
\end{abstract}


\section{Introduction}
Recent developments in artificial intelligence have witnessed a rapid expansion of potential applications of complex graph and network models in lots of fields, including Finance\cite{fan2024advanced}, big data analysis \cite{zhu2021twitter, gao2016novel}, transportation system \cite{hu2023artificial, wang2024research}, image analysis \cite{DAN2024134837, 10.11648/j.ajcst.20190202.11}, healthcare and medical \cite{Shen2024Harnessing} fields. The medical and healthcare field is always known for enormous challenges in data acquisition\cite{songgoing2023}, data quality \cite{sun2024datadrivenfilterdesignfbp}, scenario complexity\cite{doi:10.2214/AJR.23.29139}, confounding effects, and hard relationship extraction and validation\cite{cancers13133218}. The application of new concepts in graph \cite{peng2024lingcn}, transformer, neural networks \cite{zhang2024cunetunetarchitectureefficient, jiang2024advanced, zheng2024identificationprognosticbiomarkersstage} therefore has significantly reformatted the research field and enable more thorough analysis on the limited data we have.

Our paper focuses on the effect of smoking on the human brain. Smoking, especially heavy and long-term smoking, is shown harmful to cardiac, pulmonary, and vascular systems. Meanwhile, recent studies on neuro-imaging also suggested an adverse effect on the brain functions \cite{swan} and degradation in neural connections \cite{durazzo} for long-time smokers. Multiple brain regions throughout the whole brain have been clinically identified with modifications in heavy smokers, and showed a positive correlation between cigarette addiction in epidemiology studies \cite{lin2013}. 

However, the specific topological connection changes and the specific pathological pathways among different brain functional regions are yet unclear. Therefore, in this project, we aimed to apply graphical models, in specific, the Graphical Lasso (glasso) algorithm to the brain fMRI data of smokers and non-smokers to identify the Markov Random Field model (MRF) of brain connectivity of each respectively, and compare the node-wide difference between the graphs of smokers and non-smokers.

\section{Data}

Our study analyzed resting-state functional magnetic resonance imaging (rs-fMRI) data from 37 heavy smokers and 36 non-smoking control subjects. The fMRI signals were divided into 116 anatomical regions of interest and 200 volumes (each last for 2 seconds) \cite{lin2015}. To reduce auto-correlation, we selected every third volume of the data, resulting in 66 samples per subject, yielding 2,442 samples for smokers and 2,376 for non-smokers. The data were pre-processed and provided ready for analysis by Lin \cite{lin2015}.

\section{Method}

\subsection{Gaussian Undirected Model}

The Gaussian undirected graph is a graphical model indicating the adjacency relationship between every pair of variable nodes based on the independence status when condition on all other variables. The edges in the graph thus indicate a direct link between two variables, while eliminating the effect of indirect or multi-level associations. It is thus helpful to identify direct linkages between two brain regions and imply the connectivity of the functional regions.

We assumed the fMRI data follows a multi-variable Gaussian distribution $X\sim N(0, \Sigma)$ with the precision matrix $K=\Sigma^{-1}$. The conditional independence is thus achieved if and only if $K_{ij} = 0$. Then we can learn undirected graphical structure by estimating “structural” zeros in the precision matrix, or estimating the entire precision matrix and treating small entries as zero. 

\subsection{Graphical Lasso Estimation}

The Graphical Lasso (Glasso) algorithm estimates sparse inverse covariance matrices, identifying key connections between brain regions by applying a regularization technique that favors significant connections while ignoring weaker ones. This approach is particularly valuable for neuroimaging, where high-dimensional data makes it essential to pinpoint meaningful connections, especially when analyzing brain network changes in smokers.

In specific, the Graphical Lasso applies a sparsity penalty to the precision matrix K:

\begin{equation}
\hat{K}^{gl}=\arg min_K\{ -\log\det(K)+tr(SK)+\lambda||K||_1 \}\label{eq}
\end{equation}

Where $S$ is the sample covariance matrix, $\lambda$ is a tuning parameter to control the sparsity penalty. The $||.||_1$ is the $l_1$-form.

As a result, the graph adjacency matrix could be estimated based on the estimated precision matrix. In particular, we used the embedded glasso function in R package `huge`.

\subsection{Parameter tuning and RIC criteria}

To pick a proper regularization parameter value of $\lambda$ in the Glasso algorithm, we used the rotation information criterion (RIC) for every lambda value and picked the one with the best RIC score. One drawback of the method is the potential for under-selection. However, as consistency of neural connection results are more focused, it is still acceptable to have a relatively higher false negative rate.

\subsection{Similarity comparison}

We obtained the de-noised correlation matrix from the estimated precision matrix and compared it with the sample Pearson correlation matrix to verify if the algorithm captured the major relationships between variables.

We compared the node-wise similarity of connections between two graphical models. Using bootstrapping, we generated ten new datasets with 2,500 observations each, estimating an adjacency matrix for every dataset. Edges appearing in over 9 out of 10 graphs were considered stable and included in the final graph. This process was repeated for both smoker and non-smoker data, and overall graph similarity was measured using the Sorensen-Dice coefficient, defined as:

\begin{equation}
SD(A, B) = \frac{\sum (adjA \ne 0 \cap adjB\ne 0)}{\sum(adjA\ne0 \cup adjB\ne0)}\label{eq}
\end{equation}

The node-specific similarity score is calculated by the Jaccard Similarity Score, defined by:

\begin{equation}
J(a,b) = \frac{|a \cap b|} {|a \cup b|}\label{eq}
\end{equation}

Nodes with significant differences are filtered out if they have a smaller similarity score than the overall graph similarity score.

\subsection{Non-structural Machine Learning Models}

We compared our graphical models with non-structural machine learning models using Independent Component Analysis (ICA), Principal Component Analysis (PCA), and a penalized multivariate logistic model (GLMNET) with an elastic net penalty. The $\lambda$ parameter was tuned via ten-fold cross-validation, using AUC as the evaluation criterion. We then extracted correlations between brain regions and smoking status to identify significantly correlated regions.

\section{Results}

\subsection{Validation of Gaussian Undirected graphs}

\begin{figure}[htbp]
\centerline{\includegraphics[width=0.75\columnwidth]{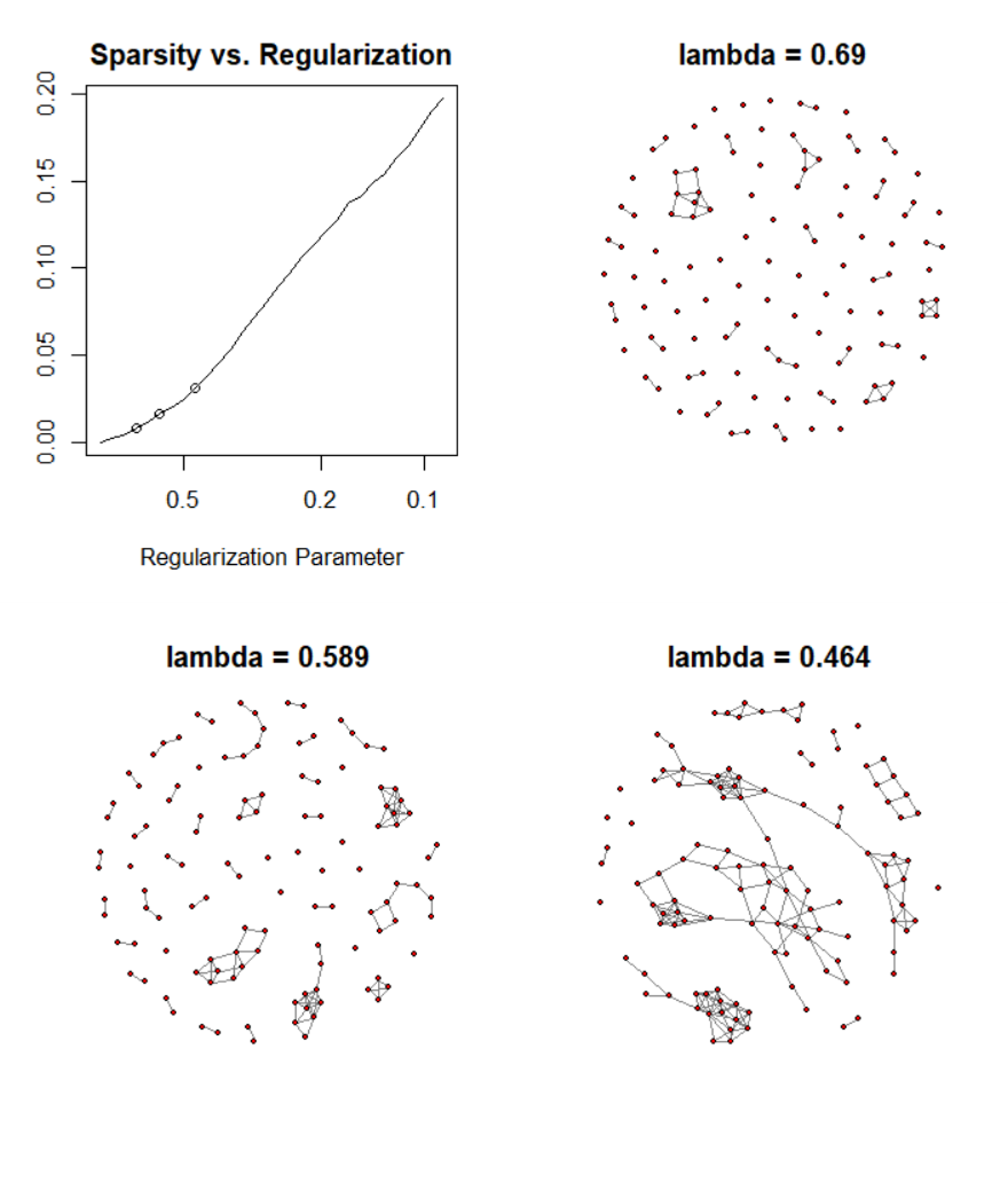}}
\caption{Estimated graph of different lambda values in non-smokers}
\label{fig1}
\end{figure}

We performed the Glasso estimation and lambda parameter tuning of the undirected graph for both smokers and non-smokers. A larger lambda would always correspond to a more sparse graph (Fig.~\ref{fig1}). Meanwhile, the optimized graphs of smokers and non-smokers showed similar results, as indicated in table ~\ref{tab1}. The similarity score of the optimized graphs for smokers and non-smokers are then calculated based on the Sorensen-Dice coefficient. 

\begin{table*}[htbp]
\caption{Comparison between the optimized graphs}
\begin{center}
\begin{tabular}{|c|c|c|c|c|c|}
\hline
& {\textbf{lambda}} & \textbf{Number of nodes}  & \textbf{Number of nodes (stable)}& {\textbf{Similarity Score$^{\mathrm{b}}$}}& \textbf{Similarity Score (stable)$^{\mathrm{a}}$} \\
\hline
\textbf{Smokers} & 0.2814 & 555 & 454 & \multirow{2}*{0.6025} & \multirow{2}*{0.631}       \\ 
\cline{1-4}
\textbf{Non-smokers} & 0.2736 & 586 & 430 &  &  \\
\hline
\multicolumn{6}{l}{$^{\mathrm{a}}$The graph-wise similarity score is calculated through Sorensen-Dice coefficient.} \\
\multicolumn{6}{l}{$^{\mathrm{b}}$Final stable graphs with common edges in \textgreater=9 bootstrap graphs throughout 10 in total}
\end{tabular}
\label{tab1}
\end{center}
\end{table*}

\begin{figure}[htbp]
\centerline{\includegraphics[width=0.7\columnwidth]{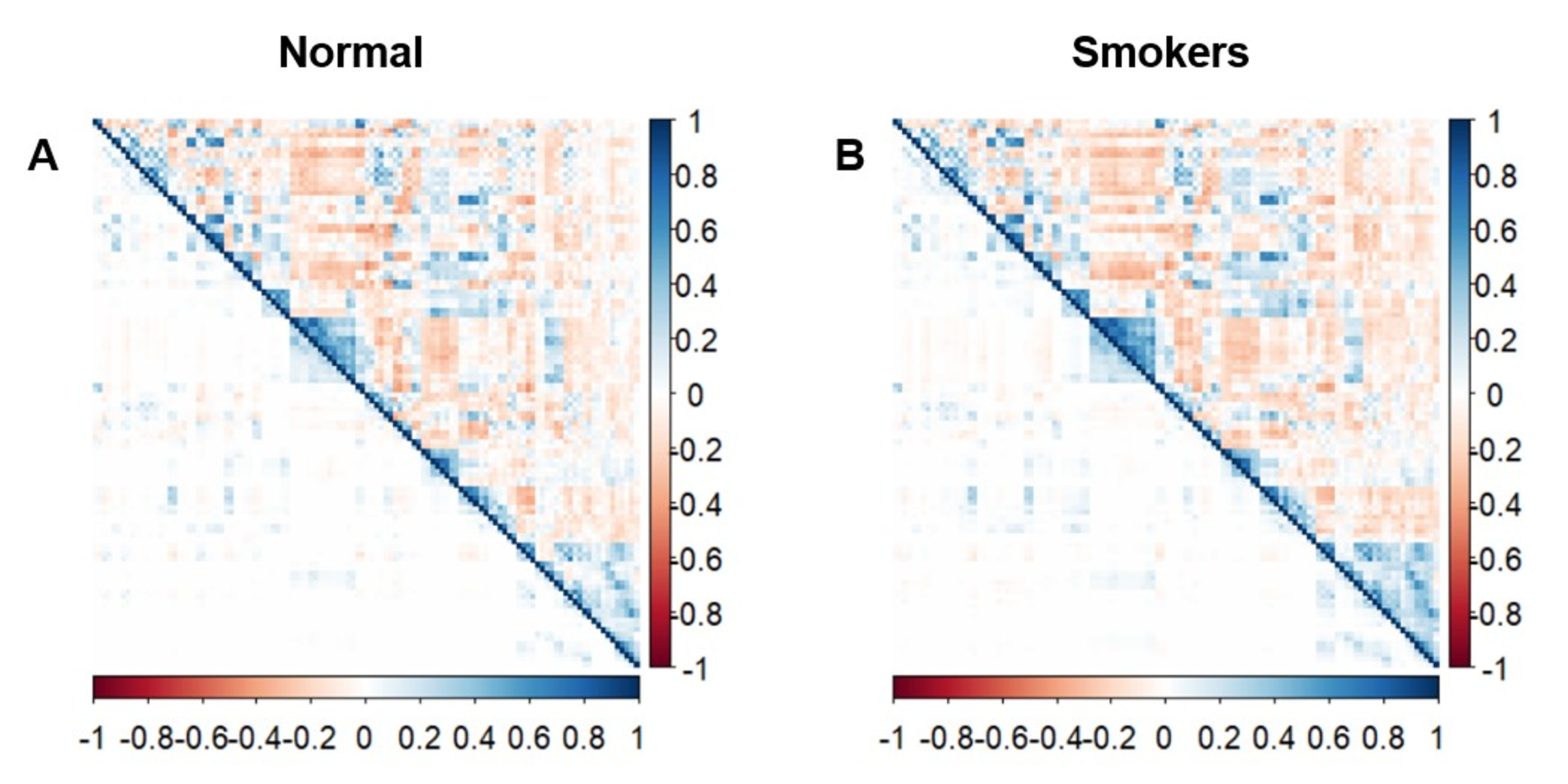}}
\caption[Connectivity between brain regions]{Connectivity between brain regions - \small Pearson correlation matrices (top triangle) and undirected graphs (bottom triangle; direct connections as discovered by Glasso, correlation calculated from estimated precision matrix). Left: non-smokers; right: smokers.}
\label{fig2}
\end{figure}

Meanwhile, we compare the reconstructed correlation matrix based on the graphic model. As indicated in Fig.~\ref{fig2},  the graphical model is able to recap most of the variance and significant correlations from the sample.

\subsection{Similarity Score Comparison based on stable edges}

After filtering out the common edges exist in more than 90 \% of the repeat graphs, we obtained an overall graph composed of common edges. The overall graph similarity score is 0.631 (Table ~\ref{tab1}), which is a little bit higher than the single graph model. Meanwhile, the number of stable edges remains at 454/555 in smokers, and 430/586 in non-smokers, indicating the graphical estimation method is relatively stable.

The node-wise Jaccard similarity score is also calculated in the stable graph and filtered with a threshold of 0.631 (Sorensen-Dice coefficient). As a result, 50 brain regions out of 116 regions are identified as significant (Table \ref{tab2}), with the smallest Jaccard Score of 0.29 in the "Temporal_Inf_L". We then take a further look at the specific connection change in the top three changed brain regions, namely “Temporal_Inf_L”, “Thalamus_R”, and “Cerebelum_Crus2”. For example, in Temporal_Inf_L, 6 connections are identified in healthy non-smokers, and 3 connections are identified in heavy smokers, while only 2 edges are shared (namely "Temporal_Mid_L", "Temporal_Inf_R"). This corresponds to the neural region functions that all belong to the Temporal region. The results also indicated loss in connection with "Frontal_Mid_Orb_L", "Frontal_Inf_Orb_L", "Parietal_Inf_L", and "Angular_L" in heavy smokers, which is worth further clinical verification (Fig.\ref{fig1}).

\begin{table}[htbp]
\caption{Top 20 regions altered between smokers and non-smokers}
\begin{center}
\begin{tabular}{|c|c|c|}
\hline
\textbf{name}& \textbf{index}& \textbf{jaccard\_score} \\
\hline
Temporal\_Inf\_L     & 8301  & 0.285714        \\
Thalamus\_R          & 7102  & 0.333333        \\
Cerebelum\_Crus2\_R  & 9012  & 0.333333        \\
Cerebelum\_10\_L     & 9081  & 0.333333        \\
Angular\_R           & 6222  & 0.35            \\
Angular\_L           & 6221  & 0.352941        \\
Precuneus\_L         & 6301  & 0.357143        \\
Caudate\_R           & 7002  & 0.375           \\
Precentral\_R        & 2002  & 0.4             \\
Frontal\_Mid\_L      & 2201  & 0.4             \\
Supp\_Motor\_Area\_R & 2402  & 0.4             \\
Cingulum\_Mid\_L     & 4011  & 0.4             \\
Frontal\_Sup\_L      & 2101  & 0.4375          \\
Occipital\_Mid\_L    & 5201  & 0.4375          \\
Calcarine\_L         & 5001  & 0.444444        \\
Parietal\_Sup\_R     & 6102  & 0.444444        \\
Precuneus\_R         & 6302  & 0.444444        \\
Frontal\_Mid\_R      & 2202  & 0.466667        \\
Temporal\_Sup\_R     & 8112  & 0.466667        \\
Frontal\_Inf\_Orb\_R & 2322  & 0.5            \\
\hline
\end{tabular}
\label{tab2}
\end{center}
\end{table}

\begin{figure}[htbp]
\centerline{\includegraphics[width=0.75\columnwidth]{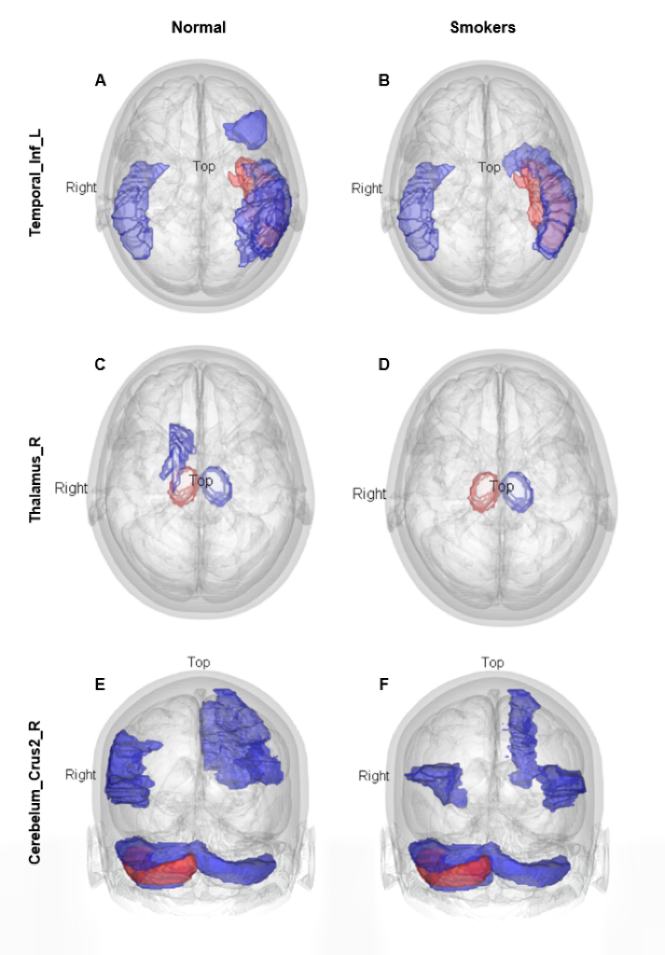}}
\caption[Brain connection visualization in Top 3 largely changed brain regions]{Brain connection visualization in Top 3 largely changed brain regions - \small Red: node region, blue: directed connected regions; left: non-smokers, right: smokers.}
\label{fig3}
\end{figure}

\subsection{Comparison between Graphical and Non-graphical Model}

We also conducted other traditional non-graphical models that are always used for fMRI data, such as Independent Component Analysis (ICA), Principal Component Analysis (PCA), and the glmnet model, and tuned with the best hyperparameters. However, traditional clustering methods such as ICA and PCA are not able to reveal the relationship alteration in brain regions, while the glmnet model does not perform well on complex fMRI data, with a maximum ROC of 0.5 (alpha=0, lambda=148.41), the same as the random effect. The extracted variable coefficient also showed no significant results ($<10^{-40}$). Therefore, we further proved that the regular non-graphical model has difficulty in identifying the difference in brain regions between smokers and non-smokers. Thus a graphical model shows more importance in complex fMRI data structure learning.

\section{Conclusion}

Our study constructed and validated Gaussian Undirected Graphs to identify significant changes in brain connectivity due to smoking. The most affected regions align with those known to be influenced by smoking \cite{wen, oneill, frank}, which offers useful guidance for further clinical research.

\section{Discussion}

\subsection{Alteration in Brain Region Connections}


Our comparison of graphical models revealed missing connections in key brain regions in heavy smokers, including the left Inferior Temporal Gyrus and the Thalamus \ref{tab2} \cite{oneill}, both of which are linked to cognitive and emotional processing \cite{wen}. These changes could contribute to the difficulty in quitting smoking by affecting decision-making and impulse control. The observed alterations in these regions highlight the potential impact of smoking on brain connectivity.

Additionally, the right posterior Crus II cerebellum (Cerebelum_Crus2_R), associated with social mentalizing and emotional experiences \cite{frank}, showed altered connectivity, though its link to smoking is less clear. Further research could explore these connections to develop targeted therapies aimed at restoring brain network integrity, potentially enhancing smoking cessation treatments.


\subsection{Model Limitations and future directions}

Although our undirected graph showed great performance on the fMRI data analysis, limitations still exist. For example, our model ignores the feature of autocorrelation in time-series data and analyzes each time stamp independently, which might result in a collinearity problem and the missing of some time-dependent features. The previous research applied a 25-second interval to reduce the collinearity, however, our dataset has a limited sample size and fMRI test length, making 25 seconds not applicable. If possible, a longer fMRI testing time or a time-series-based graphical model could be developed or applied to fMRI data in future studies.

\bibliographystyle{IEEEtran}
\bibliography{main}

\begin{thebibliography}{10}
\providecommand{\url}[1]{#1}
\csname url@samestyle\endcsname
\providecommand{\newblock}{\relax}
\providecommand{\bibinfo}[2]{#2}
\providecommand{\BIBentrySTDinterwordspacing}{\spaceskip=0pt\relax}
\providecommand{\BIBentryALTinterwordstretchfactor}{4}
\providecommand{\BIBentryALTinterwordspacing}{\spaceskip=\fontdimen2\font plus
\BIBentryALTinterwordstretchfactor\fontdimen3\font minus \fontdimen4\font\relax}
\providecommand{\BIBforeignlanguage}[2]{{%
\expandafter\ifx\csname l@#1\endcsname\relax
\typeout{** WARNING: IEEEtran.bst: No hyphenation pattern has been}%
\typeout{** loaded for the language `#1'. Using the pattern for}%
\typeout{** the default language instead.}%
\else
\language=\csname l@#1\endcsname
\fi
#2}}
\providecommand{\BIBdecl}{\relax}
\BIBdecl

\bibitem{fan2024advanced}
\BIBentryALTinterwordspacing
X.~Fan, C.~Tao, and J.~Zhao, ``Advanced stock price prediction with xlstm-based models: Improving long-term forecasting,'' \emph{Preprints}, no. 2024082109, August 2024. [Online]. Available: \url{https://doi.org/10.20944/preprints202408.2109.v1}
\BIBentrySTDinterwordspacing

\bibitem{zhu2021twitter}
W.~Zhu and T.~Hu, ``Twitter sentiment analysis of covid vaccines,'' in \emph{2021 5th International Conference on Artificial Intelligence and Virtual Reality (AIVR)}, 2021, pp. 118--122.

\bibitem{gao2016novel}
H.~Gao, H.~Wang, Z.~Feng, M.~Fu, C.~Ma, H.~Pan, B.~Xu, and N.~Li, ``A novel texture extraction method for the sedimentary structures’ classification of petroleum imaging logging,'' in \emph{Pattern Recognition: 7th Chinese Conference, CCPR 2016, Chengdu, China, November 5-7, 2016, Proceedings, Part II 7}.\hskip 1em plus 0.5em minus 0.4em\relax Springer, 2016, pp. 161--172.

\bibitem{hu2023artificial}
T.~Hu, W.~Zhu, and Y.~Yan, ``Artificial intelligence aspect of transportation analysis using large scale systems,'' in \emph{Proceedings of the 2023 6th Artificial Intelligence and Cloud Computing Conference}, 2023, pp. 54--59.

\bibitem{wang2024research}
Z.~Wang, H.~Yan, C.~Wei, J.~Wang, S.~Bo, and M.~Xiao, ``Research on autonomous driving decision-making strategies based deep reinforcement learning,'' \emph{arXiv preprint arXiv:2408.03084}, 2024.

\bibitem{DAN2024134837}
\BIBentryALTinterwordspacing
H.-C. Dan, B.~Lu, and M.~Li, ``Evaluation of asphalt pavement texture using multiview stereo reconstruction based on deep learning,'' \emph{Construction and Building Materials}, vol. 412, p. 134837, 2024. [Online]. Available: \url{https://www.sciencedirect.com/science/article/pii/S0950061823045580}
\BIBentrySTDinterwordspacing

\bibitem{10.11648/j.ajcst.20190202.11}
\BIBentryALTinterwordspacing
Y.~Song, ``Deep learning applications in the medical image recognition,'' \emph{American Journal of Computer Science and Technology}, vol.~2, no.~2, pp. 22--26, 2019. [Online]. Available: \url{https://doi.org/10.11648/j.ajcst.20190202.11}
\BIBentrySTDinterwordspacing

\bibitem{Shen2024Harnessing}
\BIBentryALTinterwordspacing
X.~Shen, Q.~Zhang, H.~Zheng, and W.~Qi, ``{Harnessing XGBoost for robust biomarker selection of obsessive-compulsive disorder (OCD) from adolescent brain cognitive development (ABCD) data},'' in \emph{Fourth International Conference on Biomedicine and Bioinformatics Engineering (ICBBE 2024)}, P.~P. Piccaluga, A.~El-Hashash, and X.~Guo, Eds., vol. 13252, International Society for Optics and Photonics.\hskip 1em plus 0.5em minus 0.4em\relax SPIE, 2024, p. 132520U. [Online]. Available: \url{https://doi.org/10.1117/12.3044221}
\BIBentrySTDinterwordspacing

\bibitem{songgoing2023}
\BIBentryALTinterwordspacing
Y.~Song, P.~Arora, R.~Singh, and et~al., ``\BIBforeignlanguage{en}{Going {Blank} {Comfortably}: {Positioning} {Monocular} {Head}-{Worn} {Displays} {When} {They} are {Inactive}},'' in \emph{\BIBforeignlanguage{en}{Proceedings of the 2023 {International} {Symposium} on {Wearable} {Computers}}}.\hskip 1em plus 0.5em minus 0.4em\relax Cancun, Quintana Roo Mexico: ACM, Oct. 2023, pp. 114--118. [Online]. Available: \url{https://dl.acm.org/doi/10.1145/3594738.3611375}
\BIBentrySTDinterwordspacing

\bibitem{sun2024datadrivenfilterdesignfbp}
\BIBentryALTinterwordspacing
Y.~Sun, L.-S. Schneider, F.~Fan, M.~Thies, M.~Gu, S.~Mei, Y.~Zhou, S.~Bayer, and A.~Maier, ``Data-driven filter design in fbp: Transforming ct reconstruction with trainable fourier series,'' 2024. [Online]. Available: \url{https://arxiv.org/abs/2401.16039}
\BIBentrySTDinterwordspacing

\bibitem{doi:10.2214/AJR.23.29139}
\BIBentryALTinterwordspacing
S.~Gupta, S.~S. Motwani, R.~H. Seitter, W.~Wang, Y.~Mu, D.~F. Chute, M.~E. Sise, D.~I. Glazer, B.~A. Rosner, and G.~C. Curhan, ``Development and validation of a risk model for predicting contrast-associated acute kidney injury in patients with cancer: Evaluation in over 46,000 ct examinations,'' \emph{American Journal of Roentgenology}, vol. 221, no.~4, pp. 486--501, 2023, pMID: 37195792. [Online]. Available: \url{https://doi.org/10.2214/AJR.23.29139}
\BIBentrySTDinterwordspacing

\bibitem{cancers13133218}
\BIBentryALTinterwordspacing
T.~K.~H. Chung, G.~Doran, T.-H. Cheung, S.-F. Yim, M.-Y. Yu, and et~al., ``Dissection of pik3ca aberration for cervical adenocarcinoma outcomes,'' \emph{Cancers}, vol.~13, no.~13, 2021. [Online]. Available: \url{https://www.mdpi.com/2072-6694/13/13/3218}
\BIBentrySTDinterwordspacing

\bibitem{peng2024lingcn}
H.~Peng, R.~Ran, Y.~Luo, J.~Zhao, S.~Huang, K.~Thorat, T.~Geng, C.~Wang, X.~Xu, W.~Wen \emph{et~al.}, ``Lingcn: Structural linearized graph convolutional network for homomorphically encrypted inference,'' \emph{Advances in Neural Information Processing Systems}, vol.~36, 2024.

\bibitem{zhang2024cunetunetarchitectureefficient}
\BIBentryALTinterwordspacing
Q.~Zhang, W.~Qi, H.~Zheng, and X.~Shen, ``Cu-net: a u-net architecture for efficient brain-tumor segmentation on brats 2019 dataset,'' 2024. [Online]. Available: \url{https://arxiv.org/abs/2406.13113}
\BIBentrySTDinterwordspacing

\bibitem{jiang2024advanced}
L.~Jiang, C.~Yu, Z.~Wu, Y.~Wang \emph{et~al.}, ``Advanced ai framework for enhanced detection and assessment of abdominal trauma: Integrating 3d segmentation with 2d cnn and rnn models,'' \emph{arXiv preprint arXiv:2407.16165}, 2024.

\bibitem{zheng2024identificationprognosticbiomarkersstage}
\BIBentryALTinterwordspacing
H.~Zheng, Q.~Zhang, Y.~Gong, Z.~Liu, and S.~Chen, ``Identification of prognostic biomarkers for stage iii non-small cell lung carcinoma in female nonsmokers using machine learning,'' 2024. [Online]. Available: \url{https://arxiv.org/abs/2408.16068}
\BIBentrySTDinterwordspacing

\bibitem{swan}
G.~Swan and C.~Lessov-Schlaggar, ``The effects of tobacco smoke and nicotine on cognition and the brain,'' \emph{Neuropsychol Rev}, p. 259–273, 2007.

\bibitem{durazzo}
T.~Durazzo, D.~Meyerhoff, and S.~Nixon, ``A comprehensive assessment of neurocognition in middle-aged chronic cigarette smokers,'' \emph{Drug Alcohol Depend}, vol. 122, p. 105–111, 2012.

\bibitem{lin2013}
F.~Lin, G.~Wu, L.~Zhu, and H.~Lei, ``Heavy smokers show abnormal microstructural integrity in the anterior corpus callosum: a diffusion tensor imaging study with tract-based spatial statistics,'' \emph{Drug Alcohol Depend}, vol. 129, p. 82–87, 2013.

\bibitem{lin2015}
\BIBentryALTinterwordspacing
------, ``Brain networks in smokers,'' \emph{Addiction Biology}, vol.~20, pp. 809--819, 2015. [Online]. Available: \url{https://doi.org/10.1111/adb.12155}
\BIBentrySTDinterwordspacing

\bibitem{wen}
\BIBentryALTinterwordspacing
M.~Wen, Z.~Yang, Y.~Wei, and et~al., ``More than just statics: Temporal dynamic changes of intrinsic brain activity in cigarette smoking,'' \emph{Addiction Biology 26( 6):e13050}, 2021. [Online]. Available: \url{https://doi.org/10.1111/adb.13050}
\BIBentrySTDinterwordspacing

\bibitem{oneill}
J.~O'Neill, M.~Tobias, M.~Hudkins, and et~al., ``Thalamic glutamate decreases with cigarette smoking,'' \emph{Psychopharmacology}, vol. 231(13), pp. 2717--24, 2014.

\bibitem{frank}
\BIBentryALTinterwordspacing
F.~V. Overwalle, Q.~Ma, and E.~Heleven, ``The posterior crus ii cerebellum is specialized for social mentalizing and emotional self-experiences: a meta-analysis,'' \emph{Social Cognitive and Affective Neuroscience}, vol.~15, p. 905–928, 2020. [Online]. Available: \url{https://doi.org/10.1093/scan/nsaa124}
\BIBentrySTDinterwordspacing

\end{thebibliography}

\end{document}